\documentclass[12pt]{article}

\usepackage[pdftex]{graphicx}

\usepackage{epsfig,cite}

\usepackage {amsmath}

\usepackage{color}
\usepackage{rotating}
\usepackage{amssymb}

\usepackage{slashed}

\include{epsf}

\newcount\timecount

\newcount\hours \newcount\minutes  \newcount\temp \newcount\pmhours

\hours = \time

\divide\hours by 60

\temp = \hours

\multiply\temp by 60

\minutes = \time

\advance\minutes by -\temp

\def\hour{\the\hours}

\def\minute{\ifnum\minutes<10 0\the\minutes

            \else\the\minutes\fi}

\def\clock{

\ifnum\hours=0 12:\minute\ AM

\else\ifnum\hours<12 \hour:\minute\ AM

      \else\ifnum\hours=12 12:\minute\ PM

            \else\ifnum\hours>12

                 \pmhours=\hours

                 \advance\pmhours by -12

                 \the\pmhours:\minute\ PM

                 \fi

            \fi

      \fi

\fi

}

\def\monthname{\relax\ifcase\month 0/\or January\or February\or

   March\or April\or May\or June\or July\or August\or September\or

   October\or November\or December\else\number\month/\fi}

\def\bold#1{\setbox0=\hbox{$#1$}%

     \kern-.025em\copy0\kern-\wd0

     \kern.05em\copy0\kern-\wd0

     \kern-.025em\raise.0433em\box0 }

\def\beq{\begin{equation}}

\def\eeq{\end{equation}}

\def\ga{\mathrel{\raise.3ex\hbox{$>$\kern-.75em\lower1ex\hbox{$\sim$}}}}

\def\la{\mathrel{\raise.3ex\hbox{$<$\kern-.75em\lower1ex\hbox{$\sim$}}}}

\def\gev{{\rm \, Ge\kern-0.125em V}}

\def\tev{{\rm \, Te\kern-0.125em V}}

\def\gyr{{\rm \, G\kern-0.125em yr}}

\def\slash#1{\rlap{\hbox{$\mskip 1 mu /$}}#1}%

\def\gappeq{\mathrel{\rlap {\raise.5ex\hbox{$>$}}

{\lower.5ex\hbox{$\sim$}}}}

\def\lappeq{\mathrel{\rlap{\raise.5ex\hbox{$<$}}

{\lower.5ex\hbox{$\sim$}}}}

\def\Toprel#1\over#2{\mathrel{\mathop{#2}\limits^{#1}}}

\def\m12{m_{1\!/2}}

\def\bea{\begin{eqnarray}}

\def\eea{\end{eqnarray}}

\def\beqar{\begin{eqnarray}}

\def\eeqar{\end{eqnarray}}

\def\m{{\cal m}}

\begin{document}

\begin{titlepage}

\pagestyle{empty}

\baselineskip=21pt


\rightline{KCL-PH-TH/2013-06, LCTS/2013-03, CERN-PH-TH/2013-017}

\vskip 0.7in

\begin{center}

{\large {\bf Associated Production Evidence against Higgs Impostors and Anomalous Couplings}}

\end{center}

\begin{center}

\vskip 0.4in

 {\bf John~Ellis}$^{1,2}$,
 {\bf Ver\'onica Sanz}$^{3,4}$
and {\bf Tevong~You}$^{1}$

\vskip 0.2in

{\small {\it

$^1${Theoretical Particle Physics and Cosmology Group, Physics Department, \\
King's College London, London WC2R 2LS, UK}\\

$^2${TH Division, Physics Department, CERN, CH-1211 Geneva 23, Switzerland}\\

$^3${Department of Physics and Astronomy, York University, Toronto, ON, Canada M3J 1P3}\\

$^4${Department of Physics and Astronomy, University of Sussex, Brighton BN1 9QH, UK}\\

}}

\vskip 0.4in

{\bf Abstract}

\end{center}

\baselineskip=18pt \noindent


{
There is still no proof that the new particle $X$ recently discovered by the ATLAS and CMS 
Collaborations indeed has spin zero and positive parity, as confidently expected. We show here that the energy dependence
of associated $W/Z + X$ production would be much less for a $J^P = 0^+$ boson with minimal couplings, 
such as the Higgs boson of the Standard Model, than for a spin-two particle with graviton-like couplings
or a spin-zero boson with non-minimal couplings. 
The $W/Z + (X \to {\bar b}b)$ signal apparently observed by the CDF and D0 Collaborations can be
used to predict the cross section for the same signal at the LHC that should be measured under the
spin-two and different spin-zero hypotheses. The spin-two prediction exceeds by an order of magnitude the upper
limits established by the ATLAS and CMS Collaborations, which are consistent with the minimal $0^+$
prediction, thereby providing {\it secunda facie} evidence against spin-two Higgs impostors.
Similar analyses of energy dependences provide
evidence against $0^-$ impostors, non-minimal scalar boson couplings, including the best LHC limits on dimension-six operators. 
Comparing the LHC vector boson
fusion cross sections at 7 and 8 TeV in the centre of mass provides additional but weaker evidence in favour of the identification
of the $X$ particle as a $J^P = 0^+$ boson with minimal couplings.}


\vfill

\leftline{
February 2013}

\end{titlepage}

\baselineskip=18pt


\section{Introduction and Summary}

The new particle $X$ with mass $\sim 125$ to 126~GeV discovered by
the ATLAS~\cite{ATLASICHEP2012} and CMS~\cite{CMSICHEP2012}
Collaborations during their searches for the Higgs boson at the LHC is
confidently expected to be a/the Higgs boson of the Standard Model. 
As such, the $X$ particle should have spin zero and positive parity~\cite{Higgs23}. However, as yet there is no convincing
evidence that it does not have spin two. Several strategies for determining the $X$
spin have been proposed, e.g., the angular distribution of $X \to \gamma \gamma$
decays~\cite{EHetal} and the kinematics of $X \to W^\pm W^{\mp*} \to \ell^+ \ell^- \nu {\bar \nu}$ and
$X \to Z Z^* \to 2 \ell^+ 2 \ell^-$ decays~\cite{EFHSYetal}. The ATLAS Collaboration has recently
released results of an analysis of the angular distribution of $X \to \gamma \gamma$
decays that favours spin zero over spin two~\cite{ATLASX2}, but not with high significance.
Little discrimination between the spin-zero and spin-two hypotheses is yet available from
analyses of $X \to W^\pm W^{\mp*}$ and $X \to Z Z^*$ decays~\cite{ATLASX2,CMSJamboree}, though the latter
do provide evidence that {\it if} it does have spin zero, it probably {\it does} have positive parity (see also~\cite{CKL}).

We have pointed out that the $V + X$
invariant mass distribution in $(V \equiv W/Z) + X$ associated production would
be entirely different under the hypotheses of a spin-two particle with graviton-like
couplings and a $0^+$ particle with a minimal scalar coupling like the Higgs 
boson of the Standard Model~\cite{EHSY}, with the $0^-$ case being intermediate.
Evidence for $V + X$ associated production at the TeVatron followed by $X \to {\bar b} b$ decay
has been provided by the CDF and D0 Collaborations~\cite{TevatronJulySearch},
but the $V + X$ invariant mass distribution has not yet been released.
We have also remarked that measurements of
the rates of $gg \to X$ production and $X \to W^\pm W^{\mp*}$, $X \to Z Z^*$ and
$\gamma \gamma$ decays disfavour simple spin-two models~\cite{ESY}, providing
{\it prima facie} evidence in favour of the spin-zero hypothesis. Also, it was
pointed out in~\cite{VBF, VBFdjouadietal} that the kinematics of vector-boson fusion (VBF) production are
different for the spin-two and -zero hypotheses, and it was suggested in~\cite{VBFdjouadietal} that the
energy dependence of VBF $X$ production could discriminate between them.

In this paper, we point out that the energy dependence of $V + X$ associated production
would also be completely different in the cases of a minimally-coupled scalar particle, a spin-two Higgs impostor with graviton-like couplings, a pseudoscalar Higgs impostor, and a scalar boson with non-minimal couplings. For example, the cross
section should grow by an order of magnitude more between the TeVatron and LHC energies
in the spin-two case than in the $0^+$ case, with the $0^-$ case being
intermediate. So far, the CMS and ATLAS Collaborations 
have only established upper limits on $V + X$ associated production followed by $X \to {\bar b} b$ decay at the LHC,
at a level somewhat larger than expected in the 
Standard Model~\cite{CMSVHbbICHEP2012,ATLASVHbbHCP2012,ATLASX2}. However, to the extent that the
TeVatron associated production signal is established, its cross section can be used to estimate
the corresponding signal strength at the LHC. The LHC cross section predicted in the spin-two
case would exceed greatly the CMS and ATLAS upper limits, whereas they are comfortably
consistent with the $0^+$ prediction. This observation provides {\it secunda facie} evidence~\cite{SF}
against the spin-two hypothesis, and also offers evidence against other possibilities
for the $J^P$ and couplings of the $X$ particle. Some weaker evidence pointing in the same direction is
provided by constraints on the energy dependence of VBF $X$ production at the LHC, as suggested in~\cite{VBFdjouadietal}.

\section{Possibilities for the $X$ Couplings to Massive Vector Bosons} 

The couplings of the Higgs-candidate $X$ to {\it massive} vector bosons are the best 
testing ground for models of electroweak symmetry breaking. In this paper we investigate the effect of 
other Lorentz structures in these couplings beyond the minimal coupling of the single $J^P = 0^+$
Higgs boson $H$ in the Standard Model:
 \bea
 {\cal L}_{0^+} \; \propto \; m_V^2 \, H V_{\mu}   V^{\mu}  
 \eea 
where $V = W^\pm, Z^0$ is a massive vector boson.
The only new operators that could appear at the dimension-five level involve an $X$ particle 
with different quantum numbers from a Standard Model Higgs, such as a pseudoscalar impostor $A$ or graviton-like couplings
of a spin-two impostor $G^{\mu \nu}$:
\bea
{\cal L}_{0^-} &= & \frac{c^{A}_V}{\Lambda} \, A \,  F_{\mu\nu} \tilde{F}^{\mu\nu}  \\
{\cal L}_{2^+} & = & \frac{c^G_{i}}{\Lambda} G^{\mu \nu} T_{\mu\nu}\ \ . 
\label{pseudoscalar}
\eea
where $T^{\mu\nu}$ is the energy-momentum
stress-tensor for a massive vector boson~\footnote{The energy dependence is dominated by the $F_{\mu \rho}  F^{\rho}_{\nu}$
term in the stress-tensor, with the contribution of the term $\propto m_V^2 V_{\mu} V_{\nu}$ being suppressed at high energies.}.
In axion-type models, $\Lambda$ is the scale of loop effects set by the decay constant of the chiral anomaly. 
In graviton-like scenarios with extra dimensions, $\Lambda \simeq {\cal O}$(TeV)
is the effective Planck mass, whereas in composite models $\Lambda \simeq M_{eff}$ would be a scale related to confinement. 
These two scenarios are, in general, related
by some suitable extension of the AdS/CFT correspondence, and we consider here the 
formulation in terms of an extra dimension~\cite{dual}.

Even if the Higgs candidate is a Lorentz scalar and a doublet under $SU(2)_L$,
there may be non-minimal dimension-six couplings with different Lorentz structures. 
Their effective Lagrangian can be written as 
\bea
{\cal L}_{d=6} \; = \; \sum_i \frac{f_i}{\Lambda^2} {\cal O}_i	\quad ,
\label{Ld6}
\eea
where here $\Lambda$ is the scale at which new physics is integrated out. 
We only need consider the subset of operators ${\cal O}_i$ which modify the $HVV$ vertices, and is not strongly constrained by electroweak precision tests. Namely, we will consider the set of operators~\cite{eduard}
\begin{eqnarray}
\label{ow}
{\cal O}_{W} & = & (D_\mu\Phi)^\dagger 
                 \widehat  W^{\mu\nu}(D_\nu\Phi) \\
\label{ob}
{\cal O}_{B} & = & 
(D_\mu\Phi)^\dagger 
     (D_\nu\Phi) \  \widehat  B^{\mu\nu} \\
\label{oww}
{\cal O}_{WW} & = & \Phi^\dagger \widehat  W^{\mu\nu} \widehat W_{\mu\nu} \Phi 
 = -\frac{g^2}{4}
(\Phi^\dagger \Phi)
W^{a\,\mu\nu}W^a_{\mu\nu}  \\
\label{obb}
{\cal O}_{BB} & = & (\Phi^\dagger \Phi) \
\widehat   B^{\mu\nu} \widehat  B_{\mu\nu}  	\quad ,
\end{eqnarray}
where $\Phi^t=(\phi_1+i \phi_2, v+H+i\phi_3)/\sqrt{2}$.
We define
\begin{equation}
\label{}
  \epsilon_i \; = \; f_i \, \frac{v^2}{\Lambda^2}	\quad ,
\end{equation}
where $i=W,B,WW,BB$,  and note that limits have been placed by global fits to LEP and LHC data in~\cite{concha} and \cite{eduard}.

A common feature of all these non-Standard Model couplings of a scalar boson to massive gauge bosons is the presence of 
derivative couplings, which leads to a non-trivial dependence on momentum, and hence on the centre-of-mass
energy.

\section{The Energy Dependence of Associated Production}

Ref.~\cite{EHSY} made the point that the kinematics of $V + X$ associated production 
at the TeVatron or the LHC would be very
different if the $X$ particle has spin two with graviton-like couplings from the case of a $J^P = 0^+$ boson
with minimal couplings like the Higgs boson of the Standard Model. Specifically, whereas in the
scalar case the $V$ and $X$ would be produced in a relative $S$ wave, the $D$ wave would
dominate in the spin-two case~\footnote{The analogous observation for associated $X$
production in $e^+ e^-$ collisions was made in~\cite{Zerwas}.}. 
As a consequence, the $V + X$ invariant mass distribution would
be peaked away from threshold in the spin-two case, whereas it is well known to be peaked
close to threshold in the scalar case. It was also observed in~\cite{EHSY}
that the case of a pseudoscalar would be intermediate, with
an invariant mass distribution that resembled more the scalar case. 
Similarly, dimension-six operators would also modify the $V + X$ mass distribution predicted in the Standard Model.  

Our observation here is that these differences in the $V + X$ kinematics lead to differences in
the centre-of-mass energy dependence of the $V + X$ associated production production cross section. To quantify this, 
we calculate the ratio of the associated production cross section at the at LHC to that at the TeVatron:
\bea
R_{AP} (X) \; = \; \frac{\sigma (p p \to V^* \to V + X, \sqrt{s})}{\sigma (p \bar{p} \to V^* \to V + X, \sqrt{s}=1.96\text{TeV})} \, .
\eea
Experiments often search for associated $V+X$ production by selecting candidates for
$Z \to e^+ e^-/\mu^+ \mu^-$ decays or $W \to e/\mu + \nu$ decays. Accordingly,
we have mimicked typical cuts to select leptonic $V$ decays in the 2-, 1- and 0-lepton subchannels. 
In the 2-lepton case, we consider the experimental cuts for $Z\to \ell^+ \ell^-$,
namely $p_{T, \ell_{1,2}}>$20 GeV and $|\eta_{\ell_{1,2}}| < 2.0$. 
In the $W \to \ell \nu$ case, we consider 1-lepton cuts with $p_T>$ 20 GeV and $|\eta_{\ell}|<$ 2.5,
and the missing transverse energy cut $\slash{E}_T>$ 25 GeV. 
In the zero-lepton channel, a basic $\slash{E}_T>$ 35 GeV cut is applied.

\begin{figure}
\includegraphics[height=5.5cm]{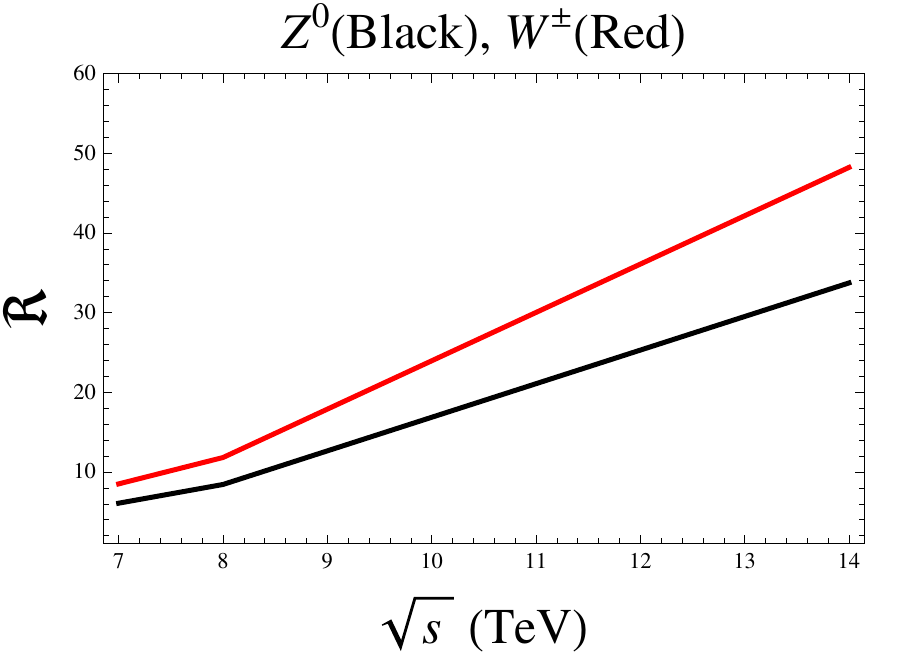}
\includegraphics[height=5.5cm]{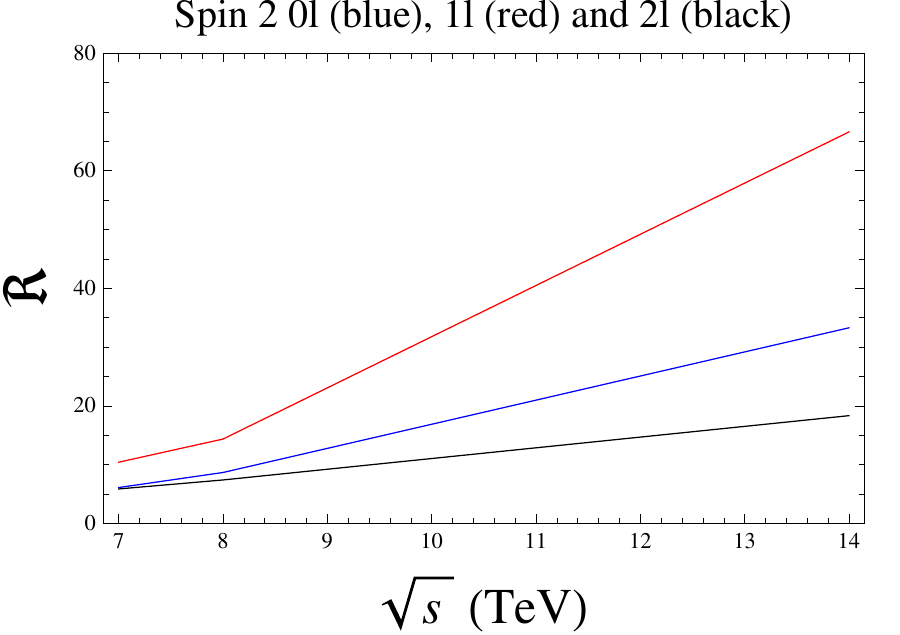}
\caption{
{\it
The energy dependences of (left) the associated production cross sections for $W^\pm + X$ (red line) and $Z^0 + X$ (black line)
and (right) the 0-, 1- and 2-lepton signals (blue, red and black lines, respectively) after experimental cuts at different LHC energies relative to the
corresponding signals at the TeVatron, as expressed via the double ratio
${\cal R}\; \equiv \;  \left(\frac{\sigma^\text{Spin~2}_\text{LHC~8}}{\sigma^\text{Spin~2}_\text{TeVatron}} 
\right) / \left(\frac{\sigma^{0^+}_\text{LHC~8}}{\sigma^{0^+}_\text{TeVatron}} \right)$.}
}
\label{fig:1l2l}
\end{figure}

The left panel of Fig.~\ref{fig:1l2l} displays the differences between the energy dependences of the cross sections
for associated production of $W^\pm + X$ (red line) and $Z^0 + X$ (black line) in the graviton-like $2^+$ and minimal $0^+$ cases
in the absence of any experimental cuts, as represented by the double ratio
\begin{equation}
{\cal R} \; \equiv \; 
\left(\frac{\sigma^\text{Spin~2}_\text{LHC}}{\sigma^\text{Spin~2}_\text{TeVatron}} \right) / \left(\frac{\sigma^{0^+}_\text{LHC}}{\sigma^{0^+}_\text{TeVatron}} \right)
\label{1l2lratio}
\end{equation}
at different LHC energies. The right panel of Fig.~\ref{fig:1l2l} displays the
corresponding double ratios after applying the experimental cuts in the 0-, 1- and 2-lepton channels (blue, red and black lines,
respectively). 

For our purposes, the relevant observation is that, both before and after applying the experimental cuts,
the energy dependences of both the 0- and 1-lepton signals are steeper than that of the 2-lepton signal. 
Both the TeVatron and the LHC experiments look for signatures that are a combination of the 0-, 1-
and 2-lepton signals. Fig.~\ref{fig:1l2l} tells us that a conservative lower bound on the energy
dependence between the TeVatron and the LHC is provided by that of the 2-lepton signal, and we concentrate on this in the following.

\begin{figure}
\begin{center}
\includegraphics[height=7cm]{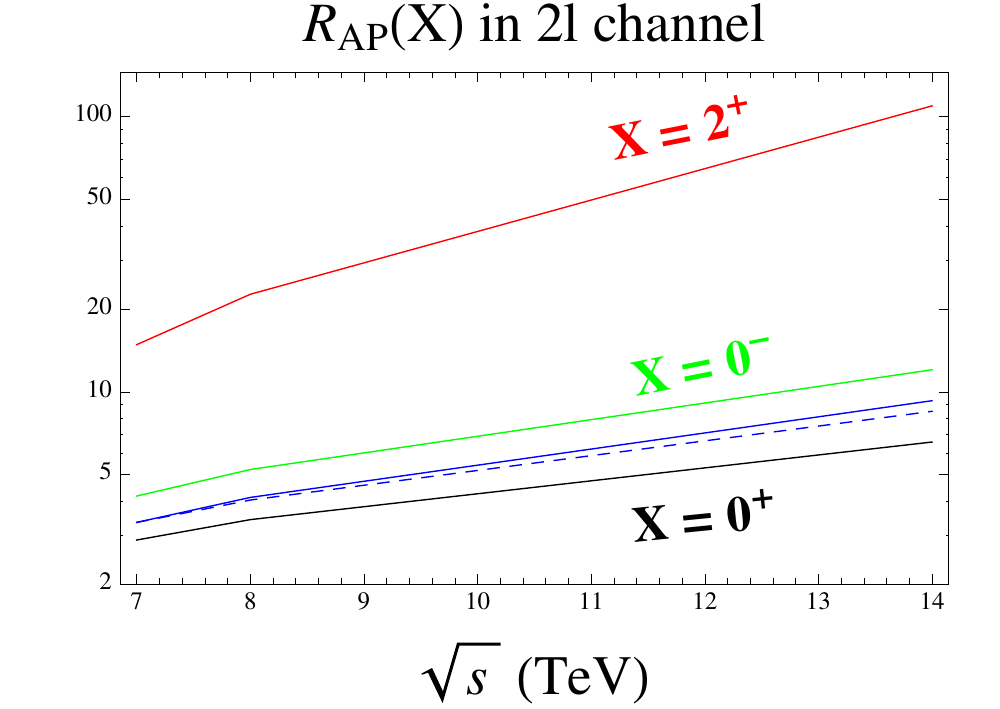}
\end{center}
\caption{{\it
The energy dependence of the cross section at the LHC relative to the cross section at the 
TeVatron for production of $X$ in association with a $Z$ boson decaying via the 2-lepton channel,
under different hypotheses for the $X$ particle: $0^+$ with minimal coupling (black), 
$0^+$ with $\epsilon_{W}=1$ (blue), $0^+$ with $\epsilon_{WW}=1$ (blue-dashed),  $2^+$ (red) and $0^-$ (green).}}
\label{fig:Ranomalous}
\end{figure}

Fig.~\ref{fig:Ranomalous}  displays the energy dependences of $(Z^0 \to \ell^+ \ell^-) + X$ 
2-lepton signal under the hypotheses that $X$ is a
minimally-coupled spin-zero boson (black line), a scalar with sizeable dimension-six operators (blue and dashed-blue),
a graviton-like spin-two boson
(red line) and a pseudoscalar (green line).
Note the logarithmic vertical scale! We see that the steepest energy
dependence is in the spin-two case, with the $0^-$ case rising more rapidly with energy
than the $0^+$ case. The effect of dimension-six operators depends on the type of operator, with 
steeper behavior for the operator with derivatives acting on the scalar boson field.

The centre-of-mass energies of immediate experimental interest to compare with 1.96~TeV
at the TeVatron are 7,8 and 14~TeV at the LHC. Accordingly, in Table~\ref{tab:Edep}
we display the ratios of the higher-energy (LHC) cross sections to that at the TeVatron
under the graviton-like spin-two, $0^-$ and $0^+$ hypotheses.
The growth of the $V + X$ associated production cross section between
the TeVatron and the LHC is much greater in the
graviton-like spin-two, pseudoscalar and non-minimal scalar cases than in the minimal scalar case. 
This suggests that a combination of TeVatron
results with relatively loose constraints on the $V + X$ associated production cross 
section at the LHC should already suffice to distinguish between these cases.

\begin{table}[t] 
\setlength{\tabcolsep}{5pt}
\center
{\renewcommand{\arraystretch}{1.3}
\begin{tabular}{|c||c|c|c|} 
\hline 
 & & &	\\[-1ex]
Model for $X$ & \large{$\frac{\rm LHC~7~TeV}{\rm Tevatron}$} & \large{$\frac{\rm LHC~8~TeV}{\rm Tevatron}$} & \large{$\frac{\rm LHC~14~TeV}{\rm Tevatron}$} \\[2ex]
\hline
Graviton-like spin $2^+$  & 16.8 & 25.1 & 119 \\\hline 
Pseudoscalar $0^-$  & 4.2 &  5.3 & 12.2 \\\hline 
Scalar $0^+$ with $\epsilon_W$=1  & 3.4 & 4.1 & 9.3 \\ \hline
Scalar $0^+$ with $\epsilon_{WW} $=1 & 3.4 & 4.0 & 8.5 \\ \hline \hline
Minimal scalar $0^+$ & 2.9 & 3.4 & 6.5 \\
\hline
\end{tabular}
}
\caption{\it The energy dependences of the cross section for $V + X$ associated production
under three $J^P$ hypotheses. The numbers shown are the ratios of the cross section
at the indicated LHC energies to the cross section at the TeVatron for the most conservative case of the 2-lepton signal.}
\label{tab:Edep} \vspace{-0.35cm}
\end{table}

\begin{figure}
\vskip 0.5in
\begin{minipage}{8in}
\hspace*{-0.7in}
\centerline{\includegraphics[height=9cm]{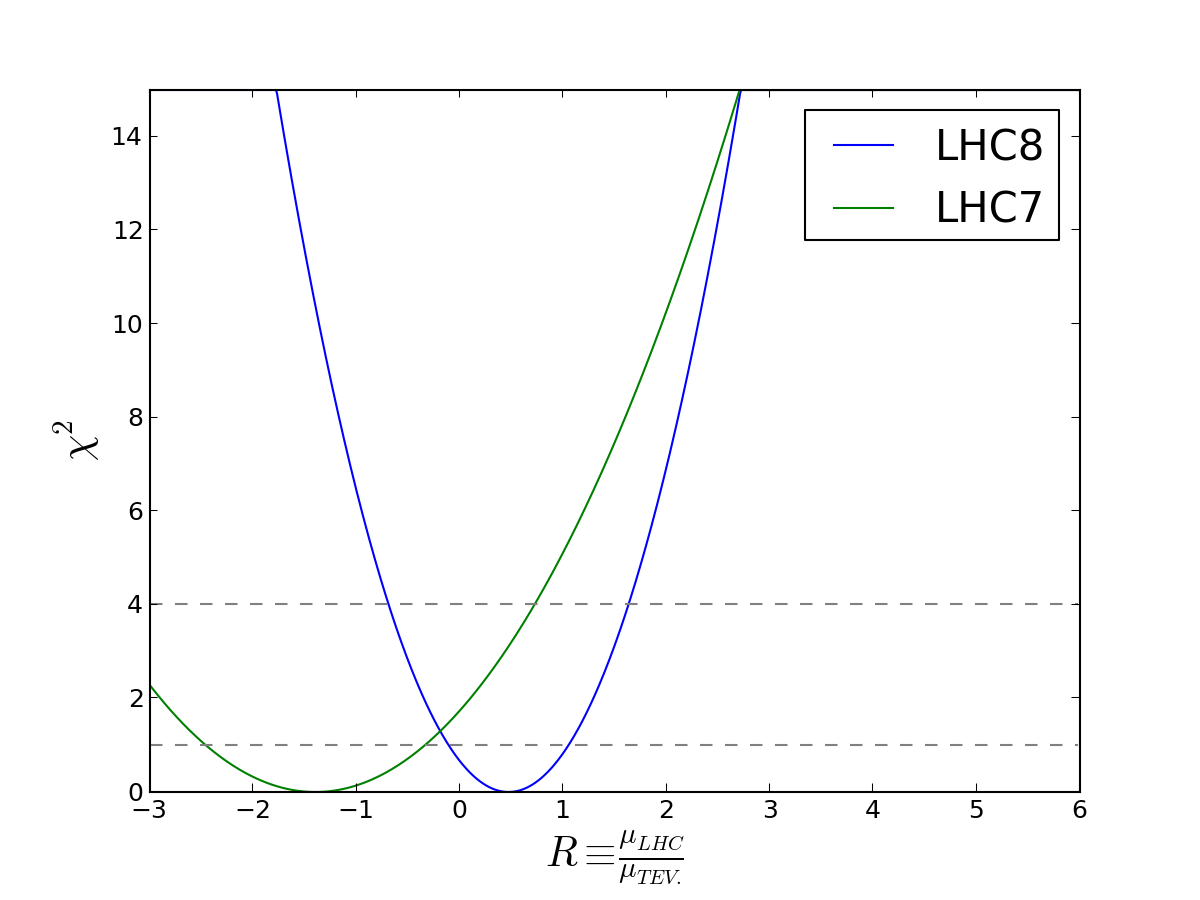}}
\hfill
\end{minipage}
\caption{
{\it
The likelihood for the ratio ${\cal R}_\text{data} = \mu_\text{LHC~8}/\mu_\text{TeVatron}$ extracted from the experimental data
at 8~TeV (blue) and 7~TeV (green). 
The most conservative spin-two expectations ${\cal R}_\text{Spin~2} = 5.9$ and $7.4$ for 7 and 8 TeV, respectively, are excluded,
and the $0^-$ expectations ${\cal R}_{0^-} = 1.48$ and $1.56$ for 7 and 8 TeV, respectively, are highly disfavoured,
whereas the $0^+$ expectation ${\cal R}_{0^+} = 1$ is quite consistent with the data.
} 
}
\label{fig:R}
\end{figure}

The TeVatron experiments CDF and D0 have reported evidence for
$V + X$ associated production with a significance between two and three $\sigma$.
Their result can be expressed as the following measurement of the strength, $\mu$,
of the signal relative to that expected for the Standard Model Higgs boson:
\begin{equation}
\mu_\text{TeVatron} \; = \; 1.56 \pm 0.73 \; ,
\label{TeVmu}
\end{equation}
whereas the CMS Collaboration reported at ICHEP2012~\cite{CMSVHbbICHEP2012} a measurement~\footnote{The more up-to-date HCP2012 results from CMS were not reported separately for 7 and 8 TeV.}:
\begin{equation}
\mu_\text{CMS~8~TeV} \; = \; 0.40 \pm 1.07 \; 
\label{CMSmu}
\end{equation}
and the ATLAS Collaboration reported at HCP2012~\cite{ATLASVHbbHCP2012} a measurement:
\begin{equation}
\mu_\text{ATLAS~8~TeV} \; = \; 1.24 \pm 1.30 \; .
\label{ATLASmu}
\end{equation}
Combining these results, we find
\begin{equation}
\mu_\text{8~TeV} \; = \; 0.74 \pm 0.84 \; .
\label{mu8}
\end{equation}
The TeVatron result (\ref{TeVmu}) and the LHC 8~TeV result (\ref{mu8})
may be combined to yield the double ratio
\begin{equation}
{\cal R}_\text{data} \; \equiv \;  \left(\frac{\sigma^\text{data}_\text{CMS~LHC~8}}{\sigma^\text{data}_\text{TeVatron}} \right) / \left(\frac{\sigma^{0^+}_\text{LHC~8}}{\sigma^{0^+}_\text{TeVatron}} \right) \;
= \; \frac{\sigma^{0^+}_\text{TeVatron}}{\sigma^\text{data}_\text{TeVatron}}\frac{\sigma^\text{data}_\text{CMS~LHC~8}}{\sigma^{0^+}_\text{LHC~8}}
=\frac{\mu_\text{LHC~8}}{\mu_\text{TeVatron}} \; = \; 0.47 \pm 0.58 \; .
\label{muratio}
\end{equation}
This is clearly quite compatible with unity, as expected for a scalar boson, but is not compatible
with the expectation for a spin-two boson with graviton-like couplings, which would be:

\begin{equation}
{\cal R}_\text{Spin~2} \; \equiv \;  \left(\frac{\sigma^\text{Spin~2}_\text{LHC~8}}{\sigma^\text{Spin~2}_\text{TeVatron}} \right) / \left(\frac{\sigma^{0^+}_\text{LHC~8}}{\sigma^{0^+}_\text{TeVatron}} \right) \;
 \simeq \; 7.4 \; 
\label{2muratio}
\end{equation}
according to the numbers in Table~\ref{tab:Edep}.
The result (\ref{muratio}) is plotted in blue in Fig.~\ref{fig:R}, together with a similar combination of the (less accurate)
7-TeV LHC data (in green). The value of the double ratio ${\cal R}_\text{Spin 2}  \simeq 5.9$ expected in that case 
(also calculated from the numbers in Table~\ref{tab:Edep}~\footnote{We recall that, as discussed earlier,
we are being conservative in basing this discussion on the energy dependence
of the 2-lepton signal, since the 1-lepton signal rises more rapidly, as seen in Fig.~\ref{fig:1l2l}.}) is also excluded.

In the $0^-$ case, the double ratio
\begin{equation}
{\cal R}_{0^-} \; \equiv \;  \left(\frac{\sigma^{0^-}_\text{LHC~8}}{\sigma^{0^-}_\text{TeVatron}} \right) / \left(\frac{\sigma^{0^+}_\text{LHC~8}}{\sigma^{0^+}_\text{TeVatron}} \right) \;
= \; \left( \frac{5.3}{3.4} \right) \; \simeq \; 1.56 \; 
\label{pseudomuratio}
\end{equation}
according to the numbers in Table~\ref{tab:Edep}, which is also inconsistent with the blue
curve in Fig.~\ref{fig:R}. Likewise, the expected ratio at 7~TeV, ${\cal R}_{0^-} \simeq 1.48$
(also as calculated conservatively from the numbers in Table~\ref{tab:Edep}), is also highly disfavoured.

Things look bad for the spin-two and $0^-$ hypotheses, with two complementary ways of excluding these 
possibilities using current data. 

As mentioned in~\cite{EHSY} and already emphasized, the $V + X$ invariant-mass distribution
in the spin-two case (and, to a lesser extent, the $0^-$ case)
would be completely different from that in the scalar case. Unfortunately, no information
is yet available from the TeVatron or LHC experiments on the $m_{V+X}$ shapes of their signals (\ref{TeVmu}).
If the shape of the TeVatron signal is {\it inconsistent} with the spin-two prediction given in~\cite{EHSY}, the spin-two hypothesis
can be excluded immediately. On the other hand, if the shape of the TeVatron signal is {\it consistent} with the spin-two prediction,
one can use the disagreement between the data (\ref{muratio}) and the spin-two prediction (\ref{2muratio}) for the ratio
of signal strengths at the TeVatron and the LHC at 8~TeV to argue that the spin-two hypothesis can be excluded in this case also.

\begin{figure}
\includegraphics[height=5.5cm]{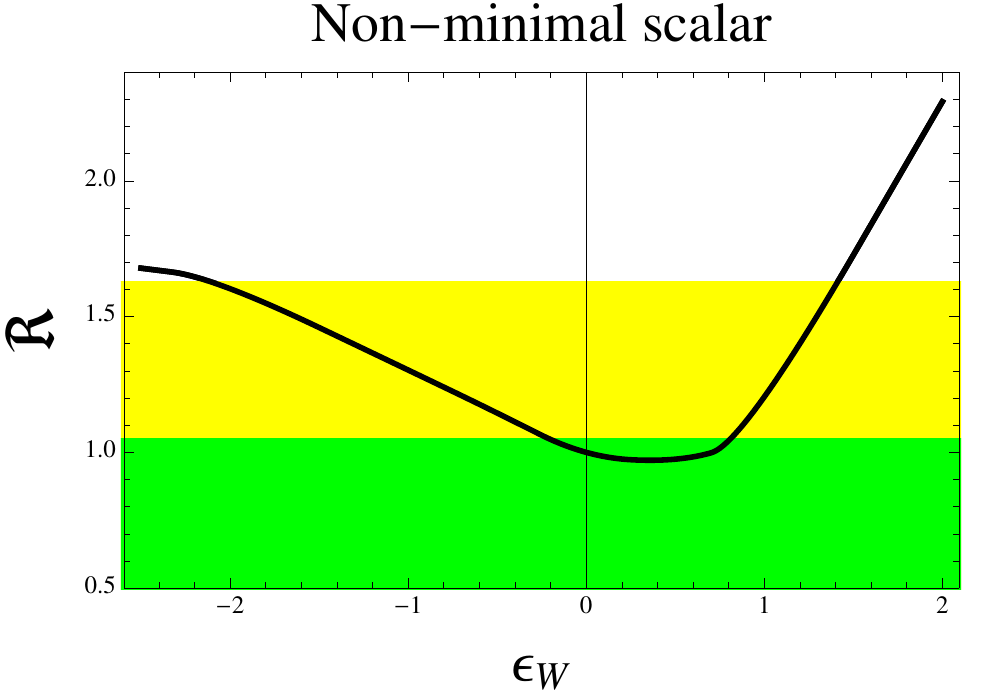}
\includegraphics[height=5.5cm]{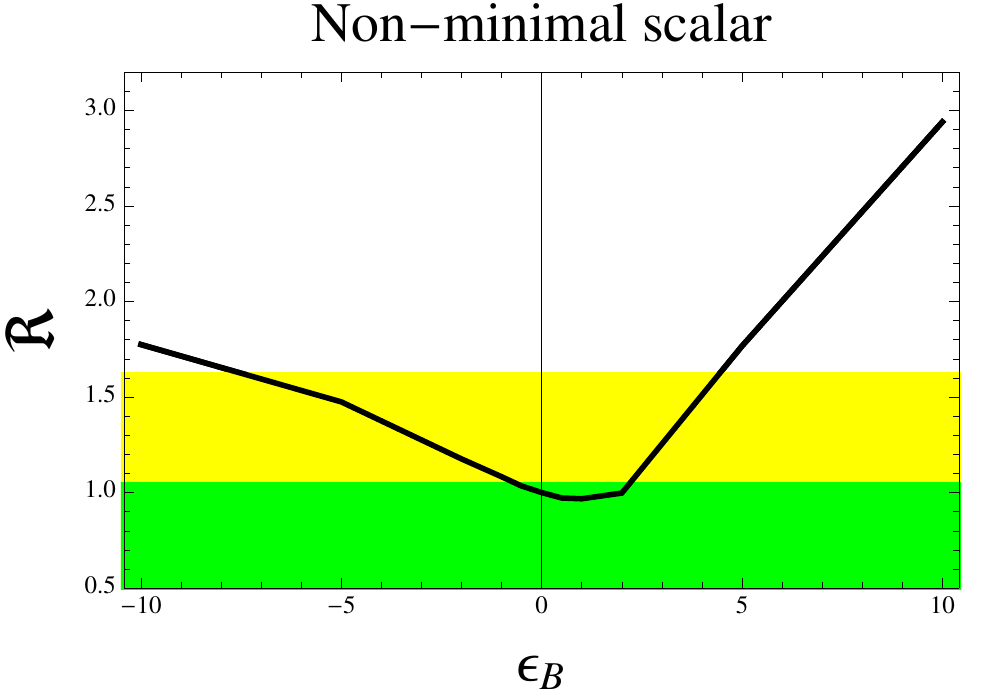}
\caption{
{\it
The effect of non-minimal couplings in the double ratio ${\cal R}$. We show the effect of $\epsilon_{W}$ (left) and $\epsilon_{B}$ (right)  in $\Re$, with the bands of 1(2)$\sigma$ in green (yellow).
} 
}
\label{eWR}
\end{figure}

We now discuss the effects of dimension-six operators on the energy dependence of $X$ in association with a
massive vector boson. Pioneering but weak bounds on $\epsilon_{W}$ and $\epsilon_B$ were derived in~\cite{eduard} 
from the LHC measurements of $X \to WW^*$ and $h\to ZZ^*$, respectively:
\bea
\textrm{\bf Higgs fit: } \epsilon_W \in [-1.3,18.5] \textrm{ and  } \epsilon_B > -9.7 \textrm{ at 95 \% CL.}	\quad .
\eea
The observable ${\cal R}$ leads to a stronger constraint on positive values of $\epsilon_W$ and $\epsilon_{B}$, 
as can be seen in Fig.~\ref{eWR}, where we plot the effects of $\epsilon_{W}$ and $\epsilon_{B}$ on ${\cal R}$, 
with the 1(2)-$\sigma$  bands shown in green (yellow). We find new combined bounds
\bea
\textrm{\bf Double ratio $\cal R$ :  } \epsilon_W \in [-1.3,1.2]  \textrm{ and } \epsilon_B \in [-7.5, 4.4]\textrm{ at 95 \% C.}	\quad
\eea
after using the experimental value of the double ratio observable. We have studied the corresponding limits on $\epsilon_{WW}$ and find them to be weaker than currently known limits coming from the signal strength of $h\to \gamma\gamma$, see Ref.~\cite{eduard}. Nevertheless, with the 14 TeV data, the energy dependence double ratios could become the best  way to precisely determine $\epsilon_{WW, BB}$ as well as $\epsilon_{W, B}$. 

 \section{The Energy Dependence of Vector Boson Fusion}

\begin{figure}
\vskip 0.5in
\begin{minipage}{8in}
\hspace*{-0.7in}
\centerline{{\includegraphics[height=7cm]{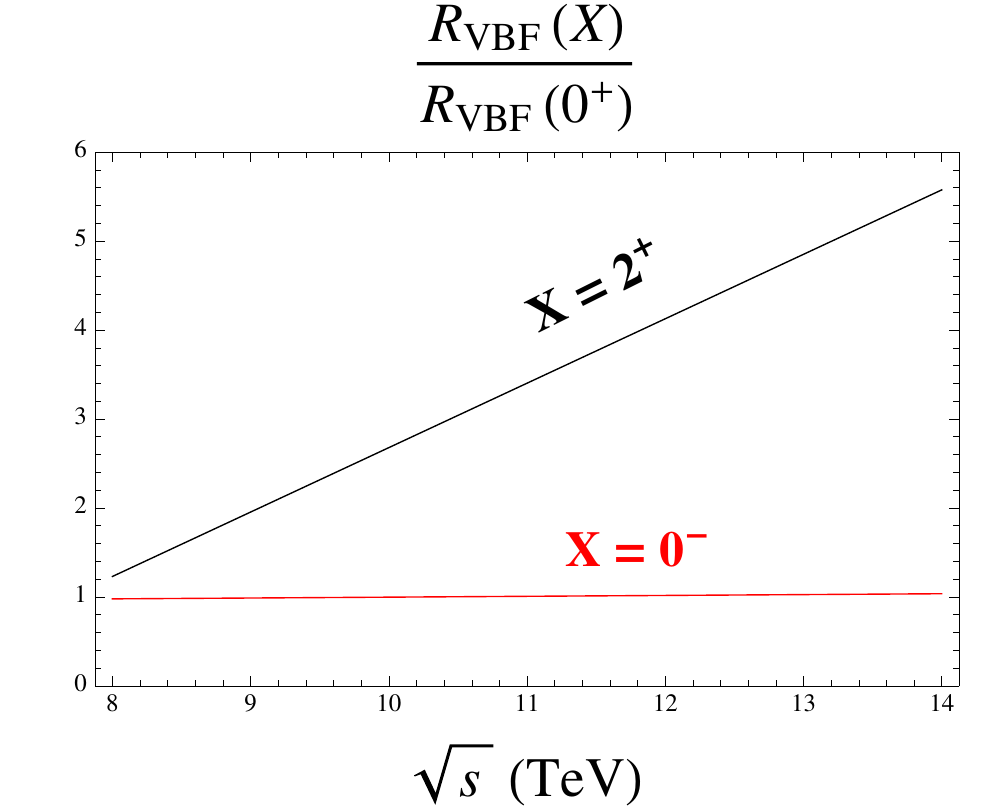}}}
\hfill
\end{minipage}
\caption{
{\it
The energy dependence of the double ratio of cross sections for VBF production of $X$,
${\cal R}_{VBF} \; \equiv \; \left(\frac{\sigma^{J^P}_{VBF}(E_\text{CM})}{\sigma^{J^P}_{VBF}(\text{8~TeV})} \right) / 
\left(\frac{\sigma^{0^+}_{VBF}(E_\text{CM})}{\sigma^{0^+}_{VBF}(8~TeV)} \right)$ for
$J^P = 2^+$ (black) and $0^-$ (red).} 
}
\label{fig:VBFpred}
\end{figure}

We have made a similar analysis of the energy dependence of the 
vector boson fusion (VBF) process. In this case, there is no
measurement from the TeVatron, so we compare production at the LHC at 7 and 8~TeV, combining ATLAS
and CMS data in each case. Fig.~\ref{fig:VBFpred} displays the predicted energy dependence of
VBF production of $X$ under different hypotheses for the $J^P$ of $X$, compared to the minimal $0^+$ case:
\begin{equation}
{\cal R}_{VBF} \; \equiv \; \left(\frac{\sigma^{J^P}_{VBF}(E_\text{CM})}{\sigma^{J^P}_{VBF}(\text{8~TeV})} \right) / 
\left(\frac{\sigma^{0^+}_{VBF}(E_\text{CM})}{\sigma^{0^+}_{VBF}(8~TeV)} \right) \, ,
\label{VBFDR}
\end{equation}
for the cases $J^P = 2^+$ (black) and $0^-$ (red). We see that the double ratio of cross sections
is unity for the $0^-$, so there is no separation power from the minimal $0^+$ in this case, 
whereas in the $2^+$ case the double ratio increases with energy by a factor 1.23 between 7~TeV
and 8~TeV, and by another factor 4.5 between 8~TeV and 14~TeV.

\begin{figure}
\vskip 0.5in
\begin{minipage}{8in}
\hspace*{-0.7in}
\centerline{\includegraphics[height=9cm]{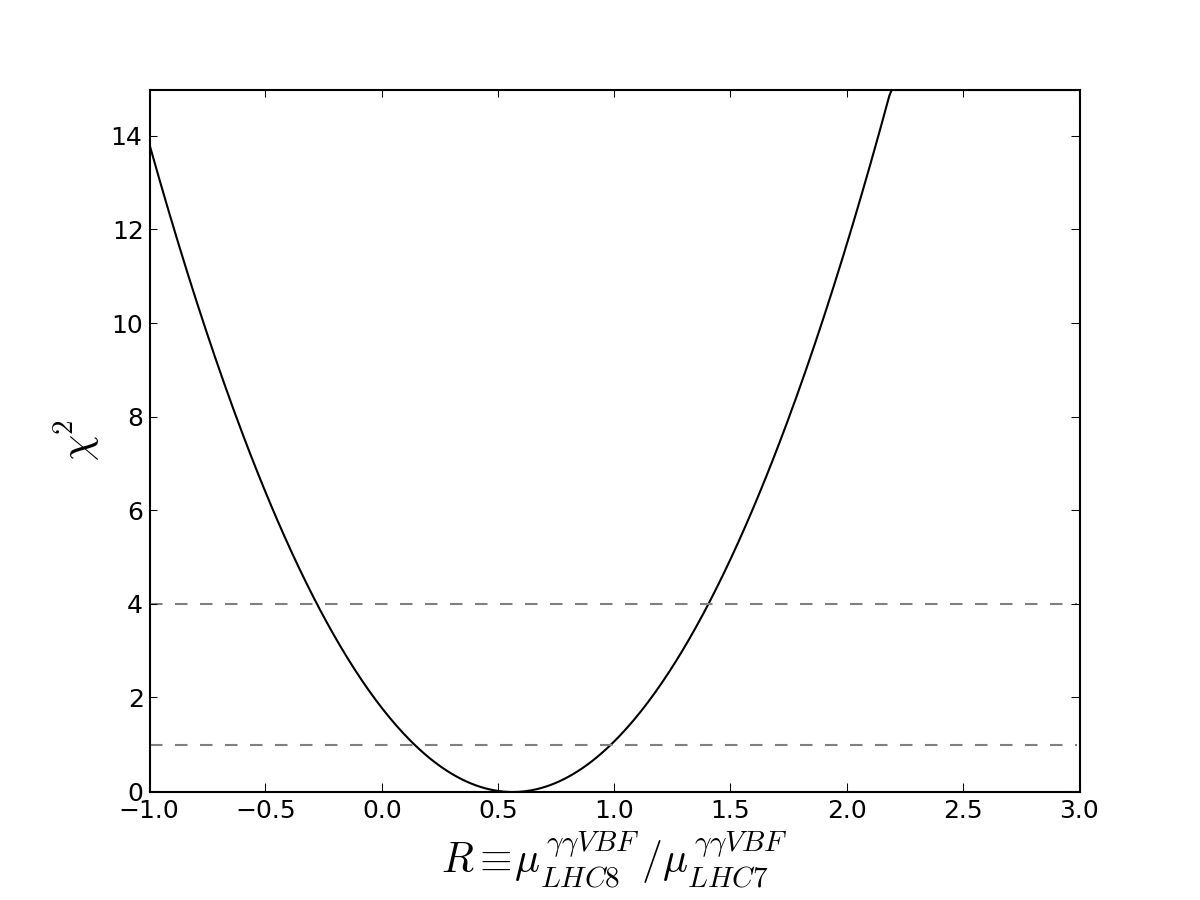}}
\hfill
\end{minipage}
\caption{
{\it
The likelihood for the ratio 
${\cal R}_\text{VBF} \equiv \mu^{\gamma\gamma \text{VBF}}_\text{LHC~8} / \mu^{\gamma\gamma \text{VBF}}_\text{LHC~7}$ extracted from a combination 
of the ATLAS and CMS searches for VBF production followed by $X \to \gamma \gamma$ decay. 
The $0^\pm$ expectations ${\cal R}_\text{VBF} = 1$ are quite compatible with the data, whereas the spin-two expectation 
${\cal R}_\text{VBF} = 1.23$ is less consistent with the data ($\Delta \chi^2 \simeq 2$).}}
\label{fig:RVBF}
\end{figure}

Fig.~\ref{fig:RVBF} shows a $\chi^2$ distribution for the ratio of VBF
signal strengths ${\cal R}_\text{VBF}$ measured at 7 and 8~TeV, {\it assuming} the rate of growth of the VBF cross section
predicted under the spin-zero hypothesis. In the spin-two case, the ratio should be $R^2_\text{VBF} = 1.23$,
and in the $0^-$ case $R^{0^-}_\text{VBF} = 1$. 
The  results shown in Fig.~\ref{fig:RVBF} are highly compatible with the prediction
${\cal R}_\text{VBF} = 1$ for $J^P = 0^\pm$, but less compatible with the spin-two prediction
($\Delta \chi^2 \simeq 2$). However, this test of the $X$ spin-parity is not yet
definitive. Similarly, limits on dimension-six anomalous couplings are weaker in VBF than in the associated production.

\section{Summary and Prospects}

We have shown in this paper that the LHC upper limits on associated $V+X$ production at 7~TeV
and 8~TeV already {\it exclude} the $J^P = 2^+$ and $0^-$ hypotheses for the $X$ particle, if one
accepts the reality of the associated production signal reported by the TeVatron experiments
CDF and D0. The LHC associated production data by themselves are not yet able to distinguish between
spin-parity hypotheses, but this may change in the future. According to the numbers in Table~\ref{tab:Edep},
between 8~TeV and 14~TeV the strength
of a spin-two signal would grow by a factor 2.12 relative faster than a $0^+$ signal, and a $0^-$ signal
would grow faster by a factor 1.51. We also used the energy dependence of associated production to 
place bounds on dimension-6 operators. These yielded stronger constraints than are currently 
obtainable from direct measurements of $h \to WW^*$ and $h \to ZZ^*$~\cite{eduard}.

In the case of the VBF process, we find that the energy dependence of the cross section
between 7~TeV and 8~TeV disfavours somwhat the spin-two hypothesis ($\Delta \chi^2 \simeq 2$)
but offers no discrimination between the $0^\pm$ hypotheses. On the other hand, the energy
dependence between 8~TeV and 14~TeV may offer prospects for the future.

The argument presented here based on the energy dependence of associated production
provides {\it secunda facie} evidence~\cite{SF} against the possibility that the $X$ particle
has spin two. Taken together with the {\it prima facie} argument based on the relative strengths of
the $X$ couplings to $\gamma \gamma, gg, W^+ W^-$ and $ZZ$ presented in~\cite{ESY}
and the $\gamma \gamma$ angular distribution presented in~\cite{ATLASX2}, it seems that
the evidence is pointing strongly towards the $0^+$ assignment for the $J^P$ of the $X$
particle with minimal couplings, in agreement with the theoretical prediction~\cite{Higgs23}.

\section*{Acknowledgements}

The work of JE was supported partly by the London
Centre for Terauniverse Studies (LCTS), using funding from the European
Research Council via the Advanced Investigator Grant 267352.
The work of TY was supported by a Graduate Teaching Assistantship from
King's College London. JE thanks CERN for kind hospitality.

\end{document}